\documentclass[10pt]{article}

\usepackage[a4paper,textwidth=18cm,textheight=24cm,top=2.85cm, bottom=2.85cm, left=1.5cm, right=1.5cm]{geometry}

\usepackage{icdp2009}

\makeatletter
\long\def\@makecaption#1#2{%
\vskip\abovecaptionskip
\sbox\@tempboxa{#1. #2}%
\ifdim \wd\@tempboxa >\hsize
#1. #2\par
\else
\global \@minipagefalse
\hb@xt@\hsize{\box\@tempboxa\hfil}%
\fi
\vskip\belowcaptionskip}
\makeatother

\usepackage{lmodern}

\usepackage{graphicx}
\usepackage{times}

\usepackage{makecell}
\usepackage{algorithm}
\usepackage{algpseudocode}
\usepackage{todonotes}

\usepackage{graphicx}
\usepackage{subfig}
\usepackage{booktabs}
\usepackage{tabularx}
\usepackage{amssymb,amsmath,amsfonts}
\usepackage{subfig}
\usepackage{tikz}
\usepackage{epic}
\usetikzlibrary{matrix}

\newtheorem{remark}{Remark}[section]

\begin{document}
\noindent

\bibliographystyle{ieeetr}

\title{Robo-Advising: Enhancing Investment with Inverse Optimization and Deep Reinforcement Learning}

\authorname{Haoran Wang  \quad  Shi Yu}
\authoraddr{\{haoran\_wang, shi\_yu\}@vanguard.com \\
The Vanguard Group, Inc.\\
Malvern, PA, USA}


\maketitle

\abstract
Machine Learning (ML) has been embraced as a powerful tool by the financial industry, with notable applications spreading in various domains including investment management. In this work, we propose a full-cycle data-driven investment robo-advising framework, consisting of two ML agents. The first agent, an inverse portfolio optimization agent, infers an investor's risk preference and expected return directly from historical allocation data using online inverse optimization. The second agent, a deep reinforcement learning (RL) agent, aggregates the inferred sequence of expected returns to formulate a new multi-period mean-variance portfolio optimization problem that can be solved using deep RL approaches. The proposed investment pipeline is applied on real market data from April 1, 2016 to February 1, 2021 and has shown to consistently outperform the S\&P 500 benchmark portfolio that represents the aggregate market optimal allocation. The outperformance may be attributed to the the multi-period planning (versus single-period planning) and the data-driven RL approach (versus classical estimation approach).

\section{Introduction}
Machine learning (ML) has been widely applied to various domains in the recent decades. Particularly in the finance domain, supervised learning, unsupervised learning and reinforcement learning, the three constituent components of ML, have led to innovative methodologies. For example,  financial time series forecast can benefit from supervised learning methods, with or without the use of deep neural networks (\cite{supervised1, supervised2}). Unsupervised learning approaches, such as clustering, have been applied to portfolio selection and optimization (\cite{unsupervised}). The recent ground-breaking achievements of reinforcement learning (RL) in games playing (e.g., AlphaGo, see \cite{AlphaGo1, AlphaGo2}) also raised significant attention to solving dynamic decision making problems in quantitative finance, including optimal trading (\cite{RL_trading1}), option hedging and pricing (\cite{RL_hedge}) and portfolio optimization (\cite{RL_portfolio1, RL_MV}).

\subsection{Challenges of applying ML to investment management}
Application of ML to the investment management is particularly promising, in view of the plentiful retail and institutional investors with increasing inflows into the asset management industry (\cite{inflow}). However, applying ML techniques to investment management has several challenges. Different from high-frequency trading where microstructure data is affluent, the available data for training ML algorithms in asset management industry is relatively limited, especially for long-term low-frequency (monthly or quarterly) rebalancing investments. Another challenge, particularly for advising retail clients, is to understand the investment objectives and the risk profiles of investors, a primary step of extreme importance. In the classical single-period mean-variance (MV) portfolio optimization setting (\cite{markowitz1952portfolio}), this amounts to learning the risk-return tradeoff parameter, also known as the risk aversion/tolerance (or equivalently, the target return, or the target variance). A good evaluation of the client's risk preference will lead to an effectively constructed portfolio that fulfills the client's risk-return tradeoff, whereas a wrongly estimated objective may lead to disastrous deviations from the client's life-long investment goals (\cite{risk_aversion1}). 

Classical procedure for estimating the risk profile is through a carefully designed questionnaire (\cite{risk_aversion2}), an approach also used by robo-advising systems (\cite{robo_advisor}). Such a procedure is usually challenging because of the ambiguity, bias, and noise in practice (\cite{questionnaire1, questionnaire2, questionnaire3}). In particular, the subjective answers clients provide in imaginary scenarios may not truly reflect their inherent risk preferences. Another more quantitatively based, systemic approach to learning clients' risk profiles is using online inverse optimization (\cite{yuetalrisk}). By directly collecting and analyzing historical portfolio holdings of a client, it is possible to infer a rational client's risk preference through solving a dual problem. This innovative approach allows continuous monitoring and evaluation of embedded portfolio risks and facilitates the automated decision process of portfolio adjustment when necessary to ensure that real portfolio risk is aligned with the investor’s true risk preference. Such a method is developed based on two fundamental methodologies: convex optimization based MV framework and learning decision making schemes through inverse optimization (IO). Clients are assumed to be rational and their portfolio decisions are near-optimal (in the sense that the portfolio is approximately on the single-period MV efficient frontier). Consequently, their decisions are affected by the risk preference factors through the MV portfolio allocation model. The inverse optimization framework then infers the risk preference factor, such as the risk aversion parameter or the expected target return, that must have been in place for those decisions to have been made. 

 Once the client's risk preference is estimated, traditional investment managers can build the portfolio that lies on the MV efficient frontier at the point corresponding to the client's risk-return tradeoff profile. Such an efficient allocation in turn depends on reliable estimates of the investment opportunity parameters, including the mean return vector and variance-covariance matrix of the underlying assets. However, in practice, the notorious difficulty of estimating the mean return has raised the third challenge for applying ML techniques to asset management. Indeed, the validity of the MV solution has been challenged for a long time (\cite{MV_estimation1, MV_estimation4}), and various econometric methods have been proposed to mitigate the effects of an inaccurate estimate of the mean return (\cite{BL, FF, robust, MV_estimation2, MV_estimation3}). In contrast to those classical estimation-based methods for solving the MV problem, the reinforcement learning (RL) field has brought completely new ideas into the MV world, particularly for the multi-period problem. Given the historical price data set, RL algorithms can learn the optimal (or near-optimal) dynamic allocation directly through trial and error in an end-to-end way, without resorting to estimating the mean return vector and the variance-covariance matrix. Combing exploration (learning) and exploitation (optimizing), RL algorithms can be flexible enough (model-free or almost model-free) to gradually find the optimal allocation strategies based on repetitive interactions with the unknown black-box investment environment. For more details on RL applications for the MV problem, we refer the interested readers to the recent works \cite{RL_MV, RL_MV2}.

\subsection{Our main contributions}
Inspired by the above two new ML approaches (IO and RL), in this paper, we propose a full-cycle, end-to-end, data-driven investment robo-advising framework, consisting of two ML agents at its core. The first IO agent collects the client's historical allocation data and solves the online inverse optimization problem to infer the risk preference. The underlying assumption is that the client's allocation is approximately on the efficient frontier of a single-period MV problem with some implicit decision parameters (i.e., risk aversion and expected return).  The IO agent's mission is to estimate the investment goal, known as portfolio-level expected return, based on historical market price and existing portfolio holdings. The second agent, an RL agent, then aggregates the time-varying portfolio-level expected returns to formulate a new multi-period MV problem and solves it using deep RL techniques. The multi-period MV optimal allocation will be the output of our full-cycle robo-advising pipeline. In practice, such a strategy can be returned back to the original client as an improved, personalized and sophisticatedly designed portfolio rebalancing strategy. 

Although our proposed robo-advising framework is mainly intended for assisting retail and/or institutional investors, due to the strict restrictions on client data disclosure in the asset management industry, we choose to work with public market portfolio data in this paper. Such a market level aggregation has been commonly adopted in finance and economics literature, including the notable Black-Litterman model (\cite{BL}), the Capital Asset Pricing Model (\cite{CAPM, CAPM1, CAPM2, CAPM3}), and the mutual fund theorem (\cite{Tobin}), to name a few. The common assumption underlying these models is that every rational market participant will choose the same allocation weights on risky assets at equilibrium, and hence the market portfolio can be seen as the overall efficient portfolio with its weights determined by the assets market values (i.e., capitalization). This paper takes the S\&P 500 portfolio as an efficient market portfolio, following the existing literature's standard practice.

We conduct empirical analysis using S\&P 500 Index data and the constituent stock price data to demonstrate the efficiency of the proposed robo-advising model. The historical S\&P 500 allocation data and market price data of constituent stocks are used by the IO agent to infer the overall market risk preference profile and expected return. For a more robust estimation, the IO agent samples the historical market allocation data at different frequencies and ensembles the estimated results to get an annualized market target return. The second agent, the RL agent, then solves the multi-period MV problem over a one-year horizon with daily rebalancing, using the annualized market return as the investment target. To ensure fair comparisons with the S\&P 500 portfolio, we design a deep RL algorithm to solve the multi-period MV problem under the no-shorting, no-borrowing constraint (i.e., allocation weights being non-negative and summing up to one). It is well known that explicit solution for such a constrained MV problem is not available to date, for either single-period, multi-period or continuous-time MV problems. We hence adopt deep neural networks as function approximators in our RL algorithm, together with a special form for the input data motivated by MV theory, to learn an efficient rebalancing strategy. It turns out, empirically, that our learned strategy consistently outperforms the S\&P 500 portfolio over the out-of-sample test data that covers the period from April 1, 2016 to February  1, 2021. It is also worth noting that, due to the aforementioned limited data challenge for training ML algorithms in asset management, the classical batch (off-line) RL training methods can easily lead to serious overfitting and produce non-robust strategies that would fail on out-of-sample data sets. This is particularly relevant in our current setting, where only the daily closing price data for the S\&P 500 stocks were used to train the deep RL algorithm. To mitigate the overfitting issue and enhance robustness, we adopt the universal training method recently proposed in \cite{RL_MV2} to train our deep RL algorithm. Such a training method additionally brings the extra benefit: even a randomly selected subset of stocks from the S\&P 500 stocks pool can have superior performance against the S\&P 500 return with high probability, as long as these stocks are allocated according to the multi-period rebalancing strategy generated by our deep RL algorithm.

The empirical findings in this work demonstrate a promising ML-based investment advising pipeline. Its success can be attributed to the two innovative and automated ML agents. Both agents make their decisions in a purely data-driven, end-to-end fashion, reducing human labor costs and decision errors.  In particular, the need to seek subjective answers from individual clients has been greatly reduced, which in turn diminishes the ambiguity and uncertainty that may arise when clients express their risk profiles to investment advisers in the classical way (e.g., through surveys or questionnaires). Moreover, by solving a multi-period investment optimization problem instead of a myopic single-period one at each time, the new investment advising pipeline has demonstrated more robust and consistent performance over a longer period in our tests. Indeed, the benefits of long-term investment have been well noted in the asset management industry. It has also been empirically observed that investors tend to have counter-cyclical risk preferences (\cite{cycle1, cycle2, cycle3}), making their investment performance more vulnerable in a volatile market.\footnote{Due to the fact that both investor's risk preference and market Sharpe ratio are counter-cyclical, solving a myopic (single-period) MV problem may lead to a withdraw of investment in risky assets, in the midst of better investment opportunity. See a more in-depth discussion in \cite{capponi} from a robo-advisor modeling perspective.} The multi-period rebalancing approach, in contrast, commits to a longer-term investment goal that is set for multiple periods ahead and, hence, greatly reduces the variability in the portfolio performance as well as the turnover rate (see Section \ref{sec:discussion}). This is supported by a further comparison between our RL based multi-period rebalancing strategy and two other approaches: a single-period MV strategy that is obtained for each quarter and a buy-and-hold strategy for the full-year horizon.

\section{Machine Learning Agents}\label{sec:model}
This section discusses the two ML agents: the inverse portfolio optimization agent and the deep reinforcement learning agent. We start with the problem formulations that the two ML agents aim to solve, followed by the general methodologies and the specific adaptations we propose.

\subsection{The inverse portfolio optimization (IPO) agent}\label{subsec:IOP}
\subsubsection{Portfolio optimization problem} 

We mention that our method is generally applicable to convex portfolio optimization models. Besides the MV model adopted in this paper, our method is also applicable to other extended formulations proposed in literature, such as portfolio selection model with transaction cost \cite{lobo2007portfolio}. Without loss of generality, we consider the Markowitz mean-variance portfolio optimization problem \cite{markowitz1952portfolio}:
	 
	\begin{align}
	\label{mean-variance portfolio}
	\tag*{PO}
    \begin{array}{llll}
         \min\limits_{\mathbf{x} \in \mathbb{R}^{n}} &  \frac{1}{2} \mathbf{x}^{T} Q \mathbf{x} - r \mathbf{c}^{T} \mathbf{x}  \\
    	\;s.t. &  A \mathbf{x} \geq  \mathbf{b},
    \end{array}
	\end{align}
	 where $Q\in\mathbb{R}^{n \times n} \succcurlyeq 0 $ is the positive semi-definite covariance matrix, $\mathbf{x}$ is portfolio allocation where each element $x_{i}$ is the holding weight of asset $i$ in the portfolio, $\mathbf{c} \in \mathbb{R}^{n}$ is the expected asset-level portfolio return, $r > 0$ is the \emph{risk tolerance} factor, $A \in \mathbb{R}^{m \times n} (n \leq m)$ is the structured constraint matrix, and $\mathbf{b}\in \mathbb{R}^{m}$ is the corresponding right-hand side in the constraints. 
	 
	 In this paper, we have $x_{i} \geq 0$ for each $i=1,2,\dots,n$, as we do not consider shorting position. In general, $\mathbf{x}$ represents the portfolio optimized for $n$ assets. Thus, $n$ is around 500 for a S\&P 500 portfolio. We further assume each element of $\mathbf{x}$ takes finite value. 
	
	 In \ref{mean-variance portfolio}, the coefficient $r$ is assigned to the linear term, and thus it represents \emph{risk tolerance} and larger $r$ indicates more preferable to profit. In literature, some formulations assign $r$ to the quadratic risk term in the objective as  \emph{risk aversion} coefficient, in which larger values lead to more conservative portfolios. Throughout the remainder of this paper, we assign $r$ to the linear term, and the learning task is to estimate \emph{risk tolerance}. In fact, the propensity for accepting risk regarding investments can be considered on a continuum \cite{Roszkowski1993}, thus \emph{risk aversion} and \emph{risk tolerance} are antonyms. 
	
	 In \ref{mean-variance portfolio}, if variables $Q, r, \mathbf{c}, A, \mathbf{b}$ are given, the optimal solution $\mathbf{x}^{*}$ can be obtained efficiently via convex optimization. In financial investment, the process is known as finding the optimal portfolio allocations, and we call it the \emph{Forward Problem}. 
	 
	 Now lets consider an \emph{Inverse Problem} where $Q, A, \mathbf{b}, \mathbf{x}^{*}$ are given, but the risk tolerance $r$ and expected return $\mathbf{c}$ are unknown. In literature,  solutions have been developed to learn $\mathbf{c}$ \cite{bertsimas2012inverse} and $r$\cite{yuetalrisk} separately via inverse optimization, assuming the other parameter is known. However, when both $r$ and $\mathbf{c}$ are unknown, to learn them simultaneously is difficult because learning the product $r \mathbf{c}^{T}$ is a non-convex problem. Thus, we propose an alternative minimization framework to learn $r$ and $\mathbf{c}$ alternatively based on the convex inverse optimization solution proposed by \cite{dong2018ioponline}.  

    \begin{remark}
    In our experiments we also explored a non-convex formulation using non-convex solver (e.g., BARON \cite{sahinidis:baron:17.8.9}) directly but was not successful. The solver was often stalled searching for solutions on some simple toy problems.    
    \end{remark}
    
    Based on our earlier work \cite{yuetalrisk}, the derivations of learning risk tolerance $r$ and expected portfolio-level return $\mathbf{c}$ are given by:

\textbf{Learning time-varying risk tolerance $r$}
	
	\begin{align}
	\label{kkt reformulation}
	\tag*{IPO-Risk}
	\begin{array}{llll}
            \min\limits_{r, \mathbf{x}, \mathbf{u}, \mathbf{z}} &\frac{1}{2}\|r - r_{t}\|^2 + \eta_t\| \mathbf{y}_{t} - \mathbf{x}\|^2\\
        	\; s.t. & A\mathbf{x}\geq \mathbf{b}, \\
        	\quad & \mathbf{u} \leq M\mathbf{z},  \\
        	\quad & A\mathbf{x} - \mathbf{b} \leq M(1-\mathbf{z}), \\
        	\quad & Q_{t}\mathbf{x} - r\mathbf{c}_{t} - A^T\mathbf{u} = 0,  \\
   	\quad & \mathbf{x} \in \mathbb{R}^{n} , \mathbf{u} \in \mathbb{R}_{+}^{m}, \mathbf{z} \in \{0, 1\}^{m}, 
	\end{array}
	\end{align}
	 where $\mathbf{z}$ is a vector of binary variables used to linearize KKT conditions, and $M$ is an appropriate number used to bound the dual variable $\mathbf{u}$ and $A\mathbf{x} - \mathbf{b}$. Clearly, \ref{kkt reformulation} is a mixed integer second order conic
	program (MISOCP), and referred to as the \emph{Inverse Portfolio Optimization of Risk Tolerance}. 

\textbf{Learning time-varying expected return $\mathbf{}$}

	\begin{align}
	\label{kkt reformulation expected return}
	\tag*{IPO-Return}
	\begin{array}{llll}
            \min\limits_{\mathbf{c}, \mathbf{x}, \mathbf{u}, \mathbf{z}} &\frac{1}{2}\|\mathbf{c} - \mathbf{c}_{t}\|^2 + \eta_t\| \mathbf{y}_{t} - \mathbf{x}\|^2\\
        	\; s.t. & A\mathbf{x}\geq \mathbf{b}, \\
        	\quad & \mathbf{u} \leq M\mathbf{z},  \\
        	\quad & A\mathbf{x} - \mathbf{b} \leq M(1-\mathbf{z}), \\
        	\quad & Q_{t}\mathbf{x} - r_{t} \mathbf{c} - A^T\mathbf{u} = 0,  \\
        	\quad & \mathbf{x} \in \mathbb{R}^{n} , \mathbf{u} \in \mathbb{R}_{+}^{m}, \mathbf{z} \in \{0, 1\}^{m}, 
	\end{array}
	\end{align}
	 and results in another MISOCP problem, and referred to as the \emph{Inverse Portfolio Optimization of Expected Return}.	 

\begin{remark}
In \ref{kkt reformulation} and \ref{kkt reformulation expected return}, $\mathbf{u}, \mathbf{z}, \eta_{t}, M$ are hyper-parameters specified respectively in the two different Inverse Problems, and they represent different variables in their own formulations. To avoid having too many symbols, we use the same variable names. 
\end{remark}

\subsubsection{Learning time-varying risk preferences and expected returns together}
Our strategy is to optimize expected returns $\mathbf{c}$ and risk preferences $r$ iteratively (the same spirit as the alternative minimization algorithm optimizing the latent variables iteratively). We combine \ref{kkt reformulation} and \ref{kkt reformulation expected return} together in a bi-level optimization procedure. In the first phase,  we learn the optimal $\mathbf{c}$ according to \ref{kkt reformulation expected return}, keeping $r$ fixed. In the second phase we learn the optimal $r$ according to \ref{kkt reformulation}, keeping the learned $\mathbf{c}$ fixed. Algorithm 1 illustrates the process of applying inverse optimization to learn $r_{t}$ and $\mathbf{c}_{t}$.

The learned risk preference $r_{t}$ and expected returns $\mathbf{c}_{t}$ represents time-varying estimations driven by a sequence of (\emph{price}, \emph{portfolio}) pairs from the start of observation till time $t$. Multiplying the learned asset-level expected return $\mathbf{c}_{t}$ with observed portfolio $\mathbf{y}_{t}$, we obtain the portfolio-level expected return $\mathbf{z}_t$, which is required for the RL agent to learn the investment strategy.   

\begin{algorithm}
	\caption{Learning Risk Tolerance and Asset-level Expected Return}
	\label{algorithm: OLF}
	\begin{flushleft}
	\textbf{Input:} (time-series portfolio and price data) $ Y_{t}, P_{t}$\\
	\textbf{Initialization:} $r_{0}$ (guess), $\mathbf{c}_{0}$ (guess), $\lambda$, $M$ (hyper-parameter)   
	\end{flushleft}
	\begin{algorithmic}[1]
		\For{$t=1$ to $T$}  
		\State receive ($Y_{t}, P_{t}$)   
		\State $Q_{t} \gets \mathbf{p}_{t}$   
		\State $\mathbf{c}_{t} \gets (Q_{t},r_{t-1}, \mathbf{y}_{t})$ Solve $\mathbf{c}_{t}$ as equation \ref{kkt reformulation expected return} 
		\State $r_{t} \gets (Q_{t},\mathbf{c}_{t}, \mathbf{y}_{t})$ Solve $r_{t}$ as equation \ref{kkt reformulation} 		
		\EndFor
	\end{algorithmic}
	\begin{flushleft}
	\textbf{Output:} Estimated $r_{t}, \mathbf{c}_{t}$ 
	\end{flushleft}
\end{algorithm}

\begin{remark}
Another approach to learn expected portfolio return directly is to formulate the mean-variance problem without explicit $r$, given by
\begin{align}
\label{mean-variance portfolio_1}
\tag*{PO-alternative}
\begin{array}{llll}
     \min\limits_{\mathbf{x} \in \mathbb{R}^{n}} &  \frac{1}{2} \mathbf{x}^{T} Q \mathbf{x}  \\
	\;s.t. &  A \mathbf{x} \geq  \mathbf{b}, \\
	       &  \mathbf{c}^{T} \mathbf{x} \geq e
\end{array}
\end{align}

We can simply learn $e$, and combine the two constraints together as:
\begin{align}
\label{newconstraints}
\tag*{PO-constraints}
\hat{A} = [c^{T}; A],  \ \ \   \hat{\mathbf{b}} = [e; \mathbf{b}] 
\end{align}

and rewrite \ref{mean-variance portfolio_1} as:
\begin{align}
\label{simple mean-variance portfolio}
\tag*{PO-b}
\begin{array}{llll}
     \min\limits_{\mathbf{x} \in \mathbb{R}^{n}} &  \frac{1}{2} \mathbf{x}^{T} Q \mathbf{x}  \\
	\;s.t. &  \hat{A} \mathbf{x} \geq  \hat{\mathbf{b}} 
\end{array}
\end{align}

Then the inverse problem becomes
\begin{align}
\label{kkt reformulation b}
\tag*{IPO-b}
\begin{array}{llll}
        \min\limits_{\hat{\mathbf{b}}, \mathbf{x}, \mathbf{u}, \mathbf{z}} &\frac{1}{2}\|\hat{\mathbf{b}} - \hat{\mathbf{b}}_{t}\|^2 + \eta_t\| \mathbf{y}_{t} - \mathbf{x}\|^2\\
    	\; s.t. & \hat{A} \mathbf{x} \geq  \hat{\mathbf{b}} , \\
    	\quad & \mathbf{u} \leq M\mathbf{z},  \\
    	\quad & \hat{A}\mathbf{x} - \hat{\mathbf{b}} \leq M(1-\mathbf{z}), \\
    	\quad & Q_{t}\mathbf{x} -  \hat{A}^T\mathbf{u} = 0,  \\
    	\quad & \mathbf{x} \in \mathbb{R}^{n} , \mathbf{u} \in \mathbb{R}_{+}^{m}, \mathbf{z} \in \{0, 1\}^{m}, 
\end{array}
\end{align}
\noindent and in $\hat{\mathbf{b}}$  we simply need to treat the first element as a learnable variable, which indicates the lowerbound of portfolio level return.

\noindent We have compared this alternative approach with Algorithm \ref{algorithm: OLF} and the results are very similar when proper upper-bounds and lower-bounds are set for $e$ as well as $\mathbf{c}$. However, formulation \ref{simple mean-variance portfolio} does not treat risk tolerance as an explicit parameter, so it is not adopted in our earlier work \cite{yuetalrisk} which focuses the problem of learning risk tolerance.   
\end{remark}

\subsection{The deep reinforcement learning (DRL) agent}\label{subsec:DRL}
\subsubsection{The multi-period MV problem}\label{classical_solution}
Once the IPO agent has estimated the time-varying expected returns $\mathcal{Z}=\{z_1, z_2, \dots, z_k\}$, $k\geq 1$, over $k$ subperiods within a year, the DRL agent can obtain the annualized return through discrete time compounding, i.e., 
$z=\Pi_{i=1}^k(1+z_i)-1$. This annualized return represents the overall risk-return profile during the period where the returns sequence $\mathcal{Z}$ was originally learned by the IPO agent. Consequently, a new multi-period MV problem can be formulated as following. Denote by $N\geq 1$ the number of rebalancing periods within a one-year horizon, and by $n\geq 1$ the number of stocks in the portfolio  to be constructed. We assume that $N\geq k$, since one of our goals is to compare the performance of the multi-period MV strategy with that of the single-period MV strategy.

Let the discrete rebalancing times be fixed at $i=0,1,2,\dots N-1$. In this paper, our focus is on daily rebalancing within a one-year horizon, therefore $N=252$. In an investment universe with $n$ stocks and one riskless asset (i.e., cash), the dynamics of the portfolio value process (i.e., the wealth process), under the self-financing condition and zero interest rate assumption for the riskless asset, follows
\begin{equation}\label{Wealth}
W_{i+1} = \sum_{j=1}^n v_i^j\frac{S_{i+1}^j}{S_i^j} + W_i-\sum_{j=1}^n v_i^j, \quad i=0,1,\dots, N-1.
\end{equation}
Here, $W_i$ represents the wealth at the $i$-th trading day, while $S_{i}^j$ represents the daily closing price of stock $j$ at date $i$. The vector $\mathbf{v}_i=(v_i^1, v_i^2, \dots, v_i^n)^T$ gives the dollar value allocations among all the $n$ stocks in the portfolio at each rebalancing day $i$. The goal of the multi-period MV problem is to find the best dynamic rebalancing strategy $\{\mathbf{v}_0$, $\mathbf{v}_1$, ..., $\mathbf{v}_{N-1}\}$ that, when implemented over the investment horizon, can achieve minimal variance while targeting a pre-specified return $z$. The precise mathematical formulation is given by
\begin{eqnarray}
& & \min_{\mathbf{v}_i, i=0,\dots, N-1} \text{Var}[W_N], \nonumber \\
& & \; s.t. \ \mathbb{E}[W_N]=1+z.\label{Multi_MV}
\end{eqnarray}
It is assumed above that the initial capital $W_0=1$. Notice that, to connect with the single-period MV problem in Section \ref{subsec:IOP}, we have used the learned and compounded return $z$ as the target in (\ref{Multi_MV}).

Given the complete knowledge of the underlying stocks mean return vectors and variance-covariance matrices at all rebalancing times $i=0,1,2,\dots, N-1$, the multi-period MV problem (\ref{Multi_MV}), as well as its variant in continuous time, have been solved rather completely with explicit solutions (see, e.g., \cite{MV_2, MV_3}). The solution technique is first to transform (\ref{Multi_MV}) into an unconstrained minimization problem using a Lagrange multiplier $2\omega$, 
\begin{equation}\label{unconstrained}
 \min_{\mathbf{v}_i, i=0,\dots, N-1}\mathbb{E}[(W_N-\omega)^2]-(\omega-1-z)^2.
\end{equation}
Such a problem can be solved based on classical dynamic programming principle (i.e., backward induction in discrete time) to obtain the optimal solution $\mathbf{v}^*_i, i=0,\dots, N-1$ that depends on the Lagrange multiplier. One can then apply the constraint $\mathbb{E}[W_N]=1+z$ to determine $\omega$ and hence the explicit optimal solution $\mathbf{v}^*_i, i=0,\dots, N-1$.

However, as noted in the Introduction, for both single-period and multi-period MV problems, it is extremely challenging to estimate the mean return vectors accurately. Moreover, the solution process involves the inversion of typically ill-conditioned variance-covariance matrices, making the allocation strategy rather sensitive to the input data. The validity of the Markowitz solution is in even greater doubt when it comes to a large amount of stocks with limited historical data, a challenging situation that is commonly referred to as the ``Markowitz's curse'' (see Section 7.4 of \cite{unsupervised}).

In the next section, we will introduce a purely data-driven, DRL-based solution to learn the optimal rebalancing strategy, without trying to estimate any of the model parameters, including the mean return vectors and variance-covariance matrices. With the help of a deep neural network approximator for the optimal rebalancing strategy, our approach can be easily extended to high dimensions, thereby devoid of the aforementioned curse of dimensionality encountered for the MV problem. Another reason for adopting deep neural networks is that MV optimal solutions under constraints are typically not in closed form. In particular, to handle the no-shorting, no-borrowing constraint, we combine a deep neural network approximation with a special input data format that is motivated by existing results for MV analysis. Our novel structure turns out to be effective when backtested on real data (see Section \ref{sec:result}).

\subsubsection{Deep reinforcement learning}
Deep reinforcement learning is a powerful technique to solve multi-period decision making problems arising in quantitative finance. Such a machine learning method has several advantages compared to classical analytic or econometric approaches, including it being model-free, data-driven, end-to-end and amenable to high-dimensional problems, among others. 


In this work, we use one of the most common RL algorithms for continuous spaces, the deep deterministic policy gradient (DDPG) algorithm. It was first introduced in \cite{DDPG} and soon became the baseline approach for solving high-dimensional RL problems with continuous states and actions. The multi-period MV problem (\ref{Multi_MV}) clearly belongs to such a class of problems. Indeed, to cast the MV problem (\ref{Multi_MV}) into the proper RL formulation, we take the ({\it wealth, allocation}) as the continuous ({\it state, action}) pair in applying DDPG. Moreover, in the original DDPG algorithm, the authors heuristically used Gaussian exploration (specifically, OU process) to balance the tradeoff between exploitation and exploration, while recent theoretical results have proved the optimality of Gaussian exploration for linear-quadratic problems with both infinite and finite horizons (\cite{Wang1, RL_MV}). The  well-known linear-quadratic structure of the MV problem (see, e.g., \cite{MV_3, RL_MV}) hence makes it more appropriate for us to work with DDPG rather than other DRL algorithms.

\subsubsection{DDPG adaptations for constrained MV problem}

Applying DDPG to solve the multi-period MV problem (\ref{Multi_MV}) is however not straightforward, as DDPG was mainly designed for classical RL problems without the variance term in the objective function. Notice that the variance term is a nonlinear function of the expectation operator, a fact that has imposed great difficulties for borrowing most RL methods into MV analysis (\cite{Shie}). Moreover, there exist additional challenges in applying RL approaches to the multi-period MV problem (\ref{Multi_MV}), or to quantitative finance problems in general. In the following, we list each of the main challenges and also provide adaptations of the original DDPG algorithm to tackle these challenges. Some of the following solutions have been applied to solve the MV problem (\ref{Multi_MV}) without constraints in \cite{RL_MV} and \cite{RL_MV2}.

\textbf{Non-classical objective.} As noted above, the variance term, which does not exist in classical RL problems, breaks the Bellman's dynamic programming principle that underlies most RL algorithms including DDPG. Nonetheless, inspired by the analytic approach in Section \ref{classical_solution}, we can use the new unconstrained criterion (\ref{unconstrained}) for DDPG. The criterion (\ref{unconstrained}) would then still contain the unknown parameter $\omega$ that corresponds to the Lagrange multiplier, making the criterion itself not fully known during learning. For that, we further adopt a separately ongoing scheme to learn $\omega$, based on the target constraint $\mathbb{E}[W_N]=1+z$ in (\ref{Multi_MV}) and stochastic approximation (\cite{SA}). The latter suggests the following updating rule to learn $\omega$,
\begin{equation}\label{w}
\omega_{k+1}=\omega_k+\alpha_k(W_N-1-z),
\end{equation}
where $\alpha_k>0$ is the learning rate. The terminal wealth $W_N$ can be replaced by a sample average of a few recently observed terminal wealth values during training to reduce fluctuations in the updating rule (\ref{w}). With this separate learning process for $\omega$, we can treat the objective (\ref{unconstrained}) as almost being stationary, by updating $\omega$ at a much slower frequency.

\textbf{Sparse, long-delayed reward.} Similar to some challenging DRL applications, including the notable AlphaGo, the current MV problem (\ref{Multi_MV}) has sparse reward. Indeed, only one terminal wealth value can be obtained for each training episode, and although this sparse signal can be backpropagated to previous time periods through the Q function update in DDPG, such a long-delayed terminal reward makes the training process less efficient and more time consuming. In order to account for this challenge, we propose to apply prioritized experience replay to DDPG rather than the original experience replay. Specifically, the new DDPG algorithm selects the terminal training experience (i.e., the terminal state-action pair) with higher probability from the replay buffer while correcting for the selection bias (\cite{PER}). Without prioritization, all the historical training outcomes would have been selected with equal probability (i.e., $1/N$), leading to  a small probability of choosing the informative signals if the number of decision making periods is large.

\textbf{Constrained actions and target tracking.} The multi-period MV problem (\ref{Multi_MV}) can take various constraints on allocation strategies. To handle the no-shorting, no-borrowing constraints   
\begin{equation}\label{constraints}
v_i^j\geq 0 \quad \text{and} \quad \sum_{j=i}^n v_i^j=W_i, \quad \text{for} \quad i=0,1, \dots, N-1,
\end{equation}
we use a softmax output layer in the actor network to automatically generate allocations. In order to better track the difference between the target return and the realized return, we also take  the actor network's input data to be $W_i-\omega_k$, rather than just $W_i$, at the $i$-th decision making time, and $\omega_k$ is the most recently learned Lagrange multiplier using (\ref{w}). This adaptation is motivated by the analytic solutions of various MV problems. Indeed, given just the no-shorting constraint (but borrowing is allowed) or given no constraints at all, the theoretical optimal allocation depends on the difference between current wealth and the true unknown $\omega$ (\cite{MV_3, MV_5, RL_MV}). Such a new form of input data also relates the current allocation decision to previously observed wealth values in a feedback loop, through the inclusion of the Lagrange multiplier.

\textbf{Limited data for training.} DRL algorithms, including DDPG, often require extensive data for training in practice, due to their typical structures consisting of multiple deep neural networks. The prevailing daily, monthly or quarterly data in investment management industry would not be sufficient to train DRL algorithms. Ad hoc training using limited data would certainly lead to serious overfitting on in-sample data, causing the learned strategy to fail when tested on real out-of-sample data (i.e., live trading). To handle this challenge, we adopt the universal training method proposed in \cite{RL_MV2}. The idea is to artificially generate randomness during training; specifically, we randomly select $n$ stocks from the S\&P 500 stocks pool to train each episode, rather than repeatedly using the same historical data for a fixed pre-selection of $n$ stocks. Such a training method has generated more robust long-short MV strategies in unforeseen market turmoils, when compared with the classical batch (off-line) RL training approach (\cite{RL_MV, RL_MV2}). Note that, with long enough training episodes, the universal training yields a rebalancing strategy that converges to the equal-weight strategy, which in itself has been proven to outperform most existing allocation strategies on out-of-sample data (\cite{equal}). However, it is also possible to obtain other strategies with early stopping, thanks to a hyper-parameters tuning process discussed below.

\textbf{Extensive hyper-parameters tuning.} Besides the sample inefficiency of most DRL algorithms, the large number of hyper-parameters to be tuned poses another challenge. The current DDPG algorithm has more than 20 hyper-parameters, including the number of hidden layers and hidden units of the two deep neural networks, training episodes, learning rates as well as initial values of different parameters, among others. To greatly reduce the workload, we distribute the DDPG training tasks over multiple CPUs via parallelization.\footnote{We trained all models on a computing platform with $36$-Core CPUs and $72$ GB RAM. The training time for a single experiment reduces from several days to around five hours.} Each DDPG model corresponds to a specific combination of hyper-parameters and the performance of all models are evaluated on the same validation data set. The best-performing DDPG model is then selected as the final model and its hyper-parameters are considered to have been best tuned. Among all the hyper-parameters, the number of training episodes determines if an early stopping is desirable for the current market period and, hence, occasionally gives rise to rebalancing strategies that outperform the equal-weight strategy generated by the longest possible training episodes.

\section{Results} \label{sec:result}

In this section, we provide details on the performance of the IPO-DRL investment strategies, generated collaboratively by the inverse portfolio optimization agent and the deep reinforcement learning agent. The data and the implementation details are delegated to the appendix, together with the description of several evaluation metrics in the results reported in this section.

First of all, the IPO agent learns the expected market target return at different frequencies and ensembles the estimates to obtain the expected annual target return. Specifically, we consider learning the quarterly and yearly market expected returns over the same period, while for the former, a simple discrete time compounding is applied to get the annualized expected return. The final ensembling step amounts to averaging the annual returns that are learned at different frequencies. Table \ref{tab:table1} shows the learned expected returns across different time scales. 

\begin{table}[ht]
\centering
\caption {\label{tab:table1} Learned expected market returns across different time periods.} 
\resizebox{0.45\textwidth}{!}{\begin{tabular}{lcccccc}
\toprule
 \multicolumn{1}{c}{\textbf{Period}} & \multicolumn{4}{c}{\textbf{Quarterly Return}}  & \textbf{Yearly Return}  & \textbf{Average} \\ 
      &  Apr.-June    & July-Sept.   & Oct.-Dec.   &  Jan.-Mar.               &                              \\    
 \cmidrule(r){1-1}  \cmidrule(r){2-5} \cmidrule(l){6-6} \cmidrule(l){7-7} 
  04-2016 to 03-2017 & $3.34\%$ & $3.33\%$ & $3.52\%$ & $3.48\%$ & $14.6\%$ & $14.5\%$\\
  04-2017 to 03-2018 & $3.55\%$ & $3.49\%$ & $3.55\%$  & $3.54\%$ & $17.6\%$ & $16.3\%$ \\
  04-2018 to 03-2019 & $3.55\%$ & $3.64\%$ & $3.51\%$ & $3.51\%$  & $18.5\%$ & $16.8\%$ \\
  04-2019 to 03-2020 & $3.57\%$ & $3.50\%$ & $3.52\%$ & $3.68\%$ & $20.9\%$ & $18.0\%$ \\
  \bottomrule
\end{tabular}}
\end{table}

\begin{figure}[ht]
	\includegraphics[width=.45\textwidth]{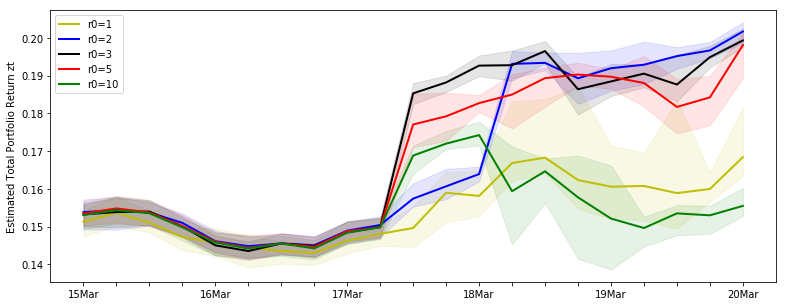}
	\caption{Time-varying annualized expected portfolio return learned by IPO agent. The estimation intervals are produced by changing the look back period $w$ of historical price. }
	\label{figure: time-varying observations}
\end{figure}

The expected portfolio returns learned by the IPO agent are illustrated in Figure \ref{figure: time-varying observations}. The time-varying expected returns $z_{t}$ are calculated as the product of the estimated expected asset-level return and the observed portfolio $\mathbf{c}_{t}^{T} \mathbf{y}_{t}$. The initial guess of risk preference $r_{0}$ has impact on the final estimation of portfolio expected return, as shown in Figure 2 by the estimation intervals of $z_{t}$ produced from five initial risk preference values ($r_{0}=1,2,3,5,10$).

The DRL agent then adopts the ensembled annual return in Table \ref{tab:table1} as the target return $z$ for the multi-period MV problem (\ref{Multi_MV}). The training and validating (model selection) of DDPG models are conducted using the two-year data before each of the one-year period in Table \ref{tab:table1}. The DRL agent then tests the selected DDPG model over these one-year periods, each of them representing an independent backtest period. 

\begin{figure*}[!t]
\subfloat[$n=5$]{\includegraphics[width=5cm]{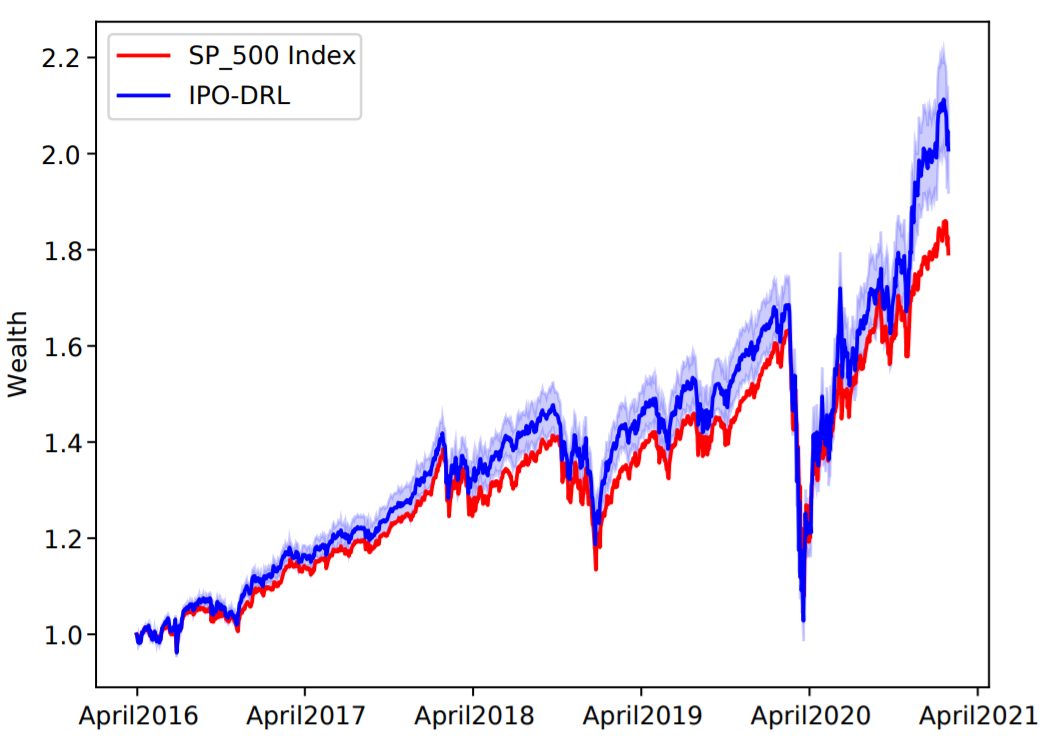}}\hfil
\subfloat[$n=10$]{\includegraphics[width=5cm]{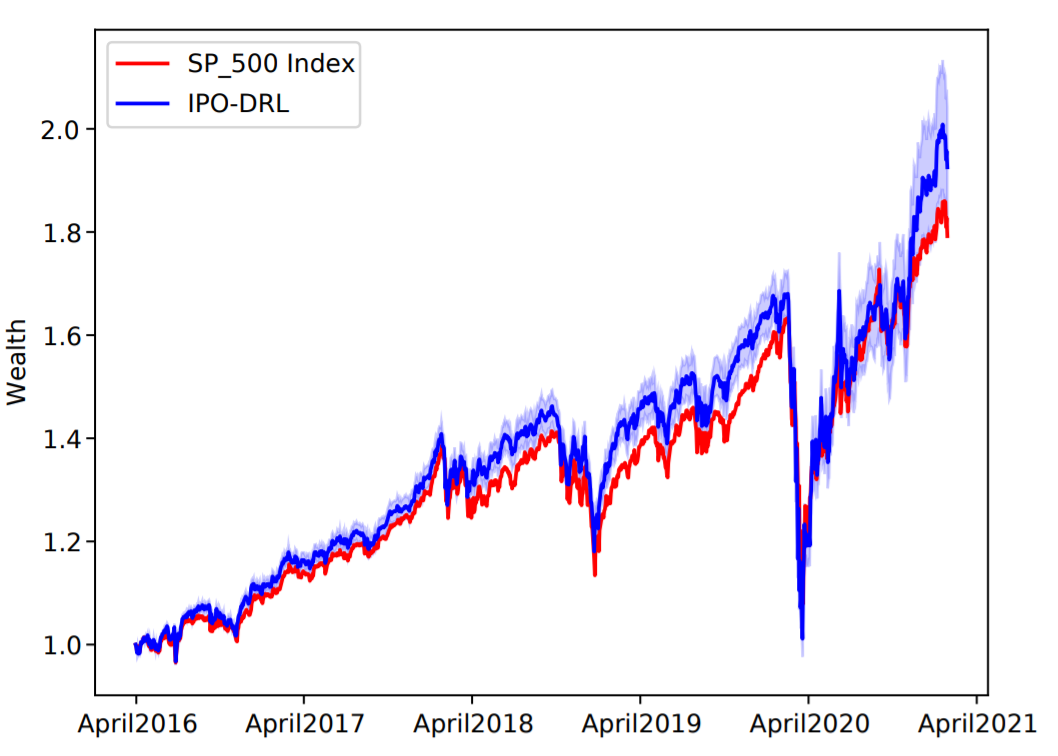}}\hfil 
\subfloat[$n=20$]{\includegraphics[width=5cm]{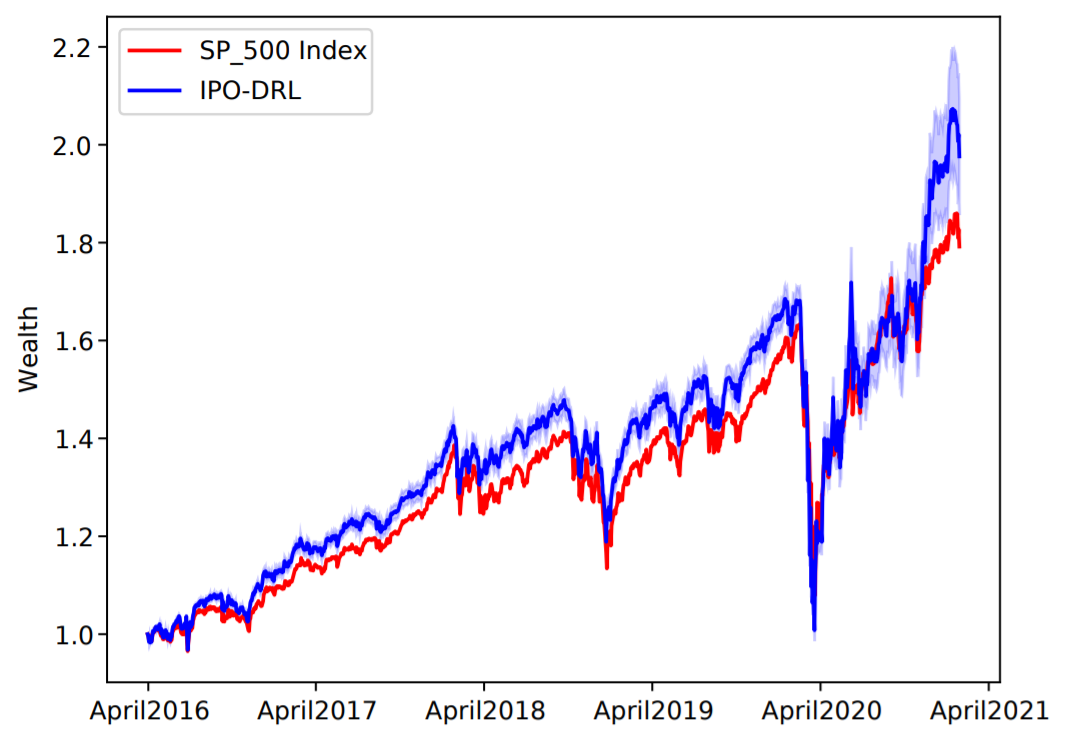}} 

\subfloat[$n=50$]{\includegraphics[width=5cm]{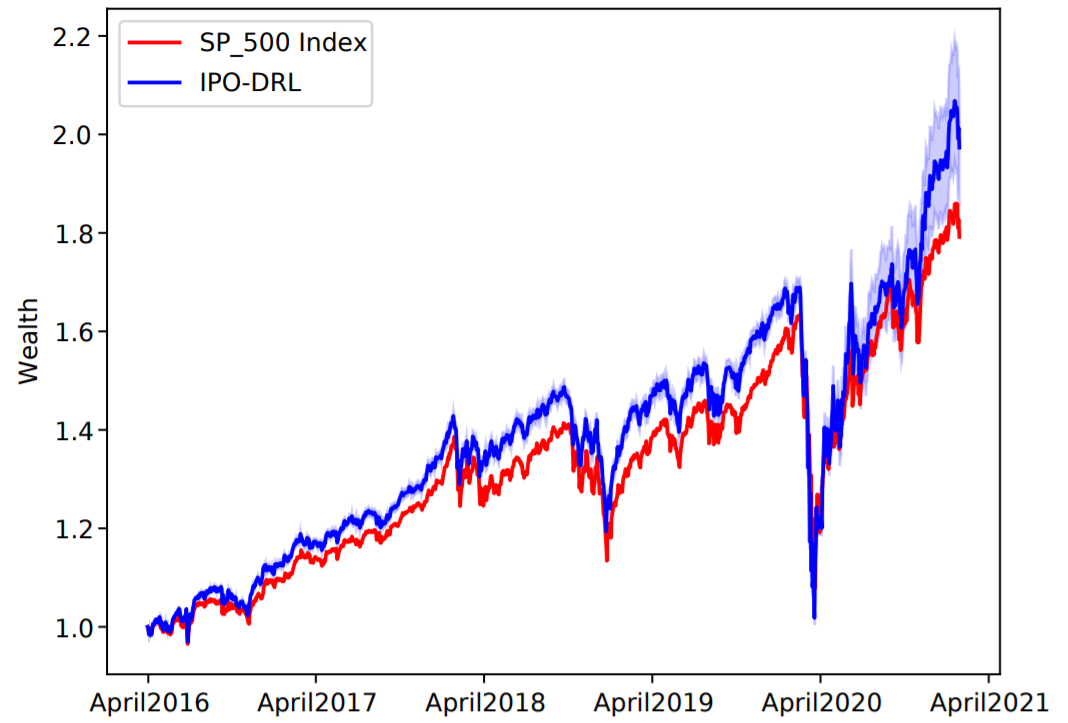}}\hfil   
\subfloat[$n=100$]{\includegraphics[width=5cm]{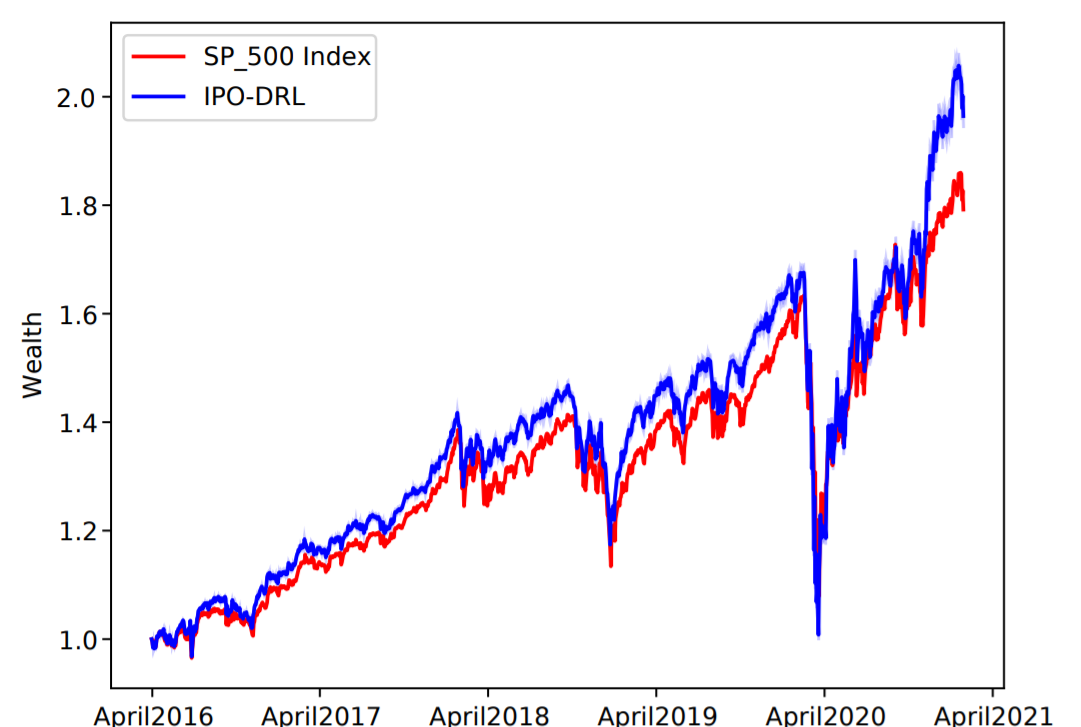}}\hfil
\subfloat[$n=200$]{\includegraphics[width=5cm]{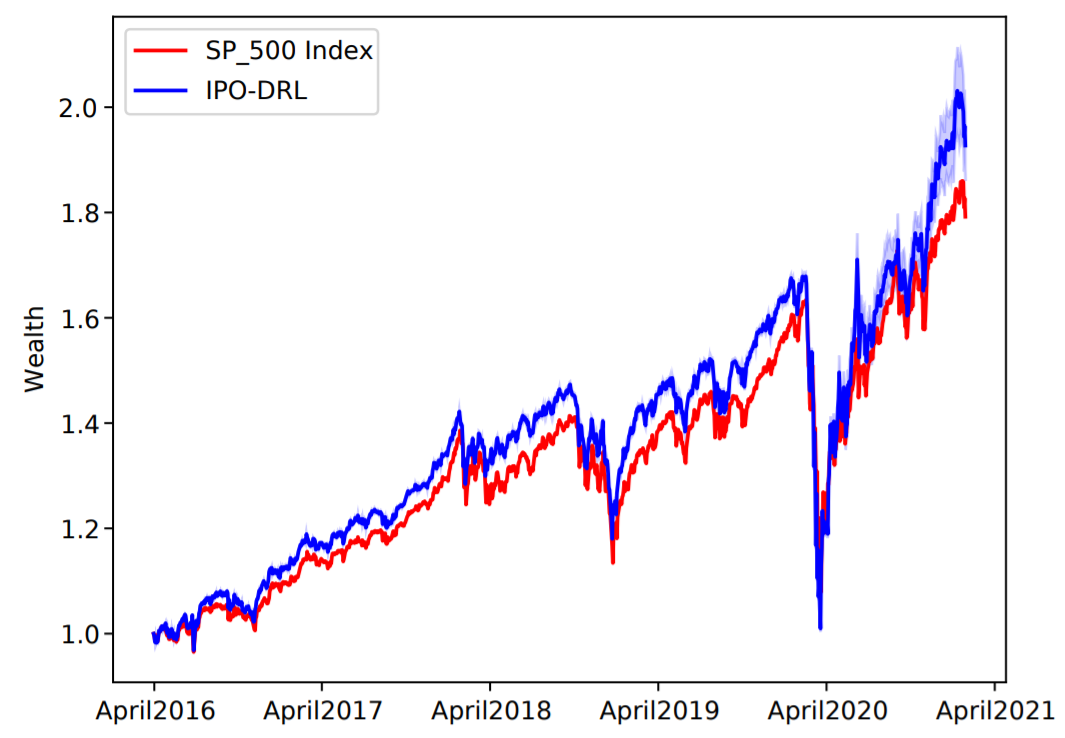}}
\caption{Test of investment strategies for different number of stocks over the period from April 1, 2016 to February 1, 2021.}\label{fig:figure1}
\end{figure*}

Figure \ref{fig:figure1} reports the backtest performance of the IPO-DRL strategy over the test period from April 1, 2016 to February 1, 2021, with one-year forward rolling windows. We also report the S\&P 500 Index as the benchmark, and two other commonly used investment strategies. The first one is the buy-and-hold strategy that only makes one-time allocation at the beginning of each one-year test period. Its allocation is computed by solving a single-period MV problem with the target being the daily return $(1+z)^{1/252}-1$, given the target annual return $z$ in the multi-period MV problem (\ref{Multi_MV}). To solve the single-period MV problem under the same no-shorting, no-borrowing constraint, we use the standard built-in optimizer SLSQP (Sequential Least SQuare Programming) of Scipy/Python. The inputs are the mean return vector and variance-covariance matrix estimated from the same two-year daily data proceeding each backtest period as in the IPO-DRL case.

The second strategy is a sequential MV problems with quarterly rebalancing. Within each one-year backtest period, we solve four single-period MV problems that are connected back to back, with the portfolio value rolling from the end of previous quarter to the start of the next. The target quarterly return is hence $(1+z)^{1/4}-1$. Similar to the buy-and-hold strategy, we impose the same no-shorting, no-borrowing constraint and apply the same SLSQP optimizer, but unlike in the previous case, we adopt the rolling-window estimation on quarterly basis. To better present the results, we choose to only plot S\&P 500 Index and the performance for the IPO-DRL strategy in Figure \ref{fig:figure1}. We also include the $95\%$ confidence band. The results for all the strategies are based on $100$ independent portfolios with random selection for the constituent stocks from the S\&P 500 pool. A summary of the performance for each strategy is presented in Table \ref{tab:table2}.

\begin{table}
\centering
\caption{\label{tab:table2}Statistics of investment strategies performance for different number of stocks over the period from April 1, 2016 to February 1, 2021.}
\resizebox{0.45\textwidth}{!}{\begin{tabular}{cccccccc} 
\toprule
\hline
\multicolumn{1}{c}{\bfseries \makecell{Number \\ of stocks}}
 & \textbf{Strategy} &  \multicolumn{1}{c}{\bfseries \makecell{Return \\ (cost-adjusted)}} & \multicolumn{1}{c}{\bfseries \makecell{Standard \\ deviation}} & \multicolumn{1}{c}{\bfseries \makecell{Sharpe \\ ratio}} & \multicolumn{1}{c}{\bfseries \makecell{Turnover\\ rate}} &  \multicolumn{1}{c}{\bfseries \makecell{Transaction \\ cost}} &  \multicolumn{1}{c}{\bfseries \makecell{Maximum \\ drawdown}}\\ 
\hline
  & IPO-DRL & $17.51\%$ & $13.48\%$ &$1.30$& $1.05\%$ & $1.21\%$ & $-41.17\%$ \\ 
$n=5$ & Buy-and-hold & $18.05\%$ & $19.38\%$ & $0.93$ & $-$  & $-$ & $-40.12\%$\\
  & Quarterly MV & $19.30\%$ & $22.12\%$ & $ 0.87$ & $89.62\%$ & $0.2\%$ & $-38.20\%$ \\ \hline
  
   & IPO-DRL & $19.71\%$ & $19.32\%$ &$1.02$& $1.07\%$ & $1.15\%$ & $-43.10\%$ \\ 
$n=10$ & Buy-and-hold & $20.98\%$ & $23.84\%$ & $0.88$ & $-$  & $-$ & $-40.68\%$\\
  & Quarterly MV & $19.95\%$ & $21.59\%$ & $ 0.92$ & $91.33\%$ & $0.3\%$ & $-40.14\%$ \\ \hline

 & IPO-DRL & $18.32\%$ & $18.97\%$ &$0.97$& $1.25\%$ & $1.06\%$ & $-41.97\%$ \\ 
$n=20$ & Buy-and-hold & $20.14\%$ & $24.08\%$ & $0.83$ & $-$  & $-$ & $-38.71\%$\\
  & Quarterly MV & $17.89\%$ & $20.62\%$ & $ 0.87$ & $98.24\%$ & $0.2\%$ & $-40.22\%$ \\ \hline
  
 & IPO-DRL & $19.25\%$ & $17.30\%$ &$1.11$& $1.13\%$ & $1.04\%$ & $-41.19\%$ \\ 
$n=50$ & Buy-and-hold & $21.68\%$ & $20.43\%$ & $1.06$ & $-$  & $-$ & $-40.99\%$\\
  & Quarterly MV & $20.69\%$ & $24.81\%$ & $ 0.83$ & $109.96\%$ & $0.3\%$ & $-39.59\%$ \\ \hline
  
  & IPO-DRL & $19.72\%$ & $18.66\%$ &$1.06$& $1.09\%$ & $1.11\%$ & $-41.97\%$ \\ 
$n=100$ & Buy-and-hold & $20.39\%$ & $21.03\%$ & $1.00$ & $-$  & $-$ & $-39.01\%$\\
  & Quarterly MV & $21.26\%$ & $22.38\%$ & $ 0.95$ & $112.18\%$ & $0.3\%$ & $-37.21\%$ \\ \hline
  
  & IPO-DRL & $18.35\%$ & $16.49\%$ &$1.11$& $1.48\%$ & $1.35\%$ & $-41.10\%$ \\ 
$n=200$ & Buy-and-hold & $22.39\%$ & $22.92\%$ & $0.97$ & $-$  & $-$ & $-38.58\%$\\
  & Quarterly MV & $19.51\%$ & $21.87\%$ & $ 0.89$ & $111.24\%$ & $0.4\%$ & $-40.11\%$ \\ \hline

\bottomrule
\end{tabular}}
\end{table}

\section{Discussions} \label{sec:discussion}
\textbf{Portfolio concentration and diversification.}\\
The IPO-DRL strategy consistently demonstrates a high level of diversification among the underlying stocks in our empirical tests. Indeed, some (but not all) finally selected IPO-DRL allocations after training and validation have converged to the equal-weight allocation, thereby effectively avoiding the concentration of allocations. In contrast, both the buy-and-hold and the quarterly MV solutions have shown the tendency to concentrate positive holdings on a small subset of stocks. For instance, in the case where $n=5$, the two classical MV strategies often generate positive holdings for two or three stocks, whereas when $n=10$, the number of invested stocks typically increases to four or five. Moreover, for the quarterly rebalanced MV problem, a high turnover rate is prevailing in our experiments (see Table \ref{tab:table2}. This is mainly due to the fact that the purchased stocks change significantly from quarter to quarter; we observe that the previously longed stocks are often liquidated at each rebalancing time while some new stocks are included in the portfolio for the current quarter . Such a higher turnover rate leads to a higher standard deviation for the classical MV solutions, but not a higher transaction cost due to the low rebalancing frequency over the whole investment horizon.

\noindent\textbf{Rebalancing frequency and performance}\\
Specifically in terms of standard deviation and Sharpe ratio, the IPO-DRL strategy consistently outperforms the other two classical MV strategies, as shown in Table \ref{tab:table2}. This may suggest that, with a multi-period MV formulation (\ref{Multi_MV}), the IPO-DRL strategy can benefit from a closer track of ongoing performance based on the current wealth and the target related Lagrange multiplier. The buy-and-hold strategy (with no intermediate rebalancing) and the  MV strategy with quarterly rebalancing may fail to reduce uncertainty of returns due to delayed responses.

Although the IPO-DRL strategy rebalances daily, the resulting turnover rate is surprisingly small for all of our experiments in Table \ref{tab:table2}. This indicates that the selected stocks remain almost unchanged during the whole investment horizon (just as the buy-and-hold strategy), while the allocation weight on each stock is only very mildly tuned from day to day. In addition, it is worth noting that, although with daily rebalancing, the IPO-DRL strategy has accrued small cumulative transaction cost throughout the investment horizon in all our experiments, demonstrating its feasibility for investment practice. 

\noindent\textbf{Data-driven learning versus classical estimation}\\
In addition to the multi-period formulation, the IPO-DRL strategy differs fundamentally from the other two classical MV strategies in that the expected returns and variance-covariance matrices do not need to be estimated {\it a priori}. Recall that the DDPG algorithm generates the IPO-DRL strategy directly based on evaluating and improving the allocation strategies using the raw price data, without any explicit estimation of model parameters. On the other hand, the other two classical MV strategies require model estimations as the input in order to generate allocation decisions. Consequently, it is shown in Table \ref{tab:table2} that the uncertainty of the estimations has led to the high turnover rates and frequent changes of selected stocks for the quarterly MV strategy which, in turn, amplifies the standard deviation of the portfolio returns.

\section{Related Works}

Our work is related to the general inverse optimization problem, in which the learner seeks to infer the missing information of the underlying decision model from observed data, assuming that the decision maker is rationally making decisions \cite{ahuja2001inverse}. Among others, \cite{dong2018ioponline} develops an online learning algorithm to infer the utility function or constraints of a decision making problem from sequentially arrived observations of (\emph{signal, decision}) pairs. Their formulation is particularly relevant to the time-varying nature of risk preferences and expected returns considered in this paper.

Also related to our work is \cite{bertsimas2012inverse}, which creates a novel reformulation of the Black-Litterman (BL) framework by using techniques from inverse optimization. There are two main differences between \cite{bertsimas2012inverse} and our approach. First, the problems we study are essentially different. \cite{bertsimas2012inverse} seeks to reformulate the BL model while we focus on learning specifically the investor's risk preferences together with expected portfolio returns. Second, \cite{bertsimas2012inverse} considers a deterministic setting in which the parameters of the BL model are fixed and uses only one portfolio as the input. In contrast, we believe that investor's risk preferences and expected returns are time-varying and propose an inverse optimization approach to learn them in an online setting with as many historical portfolios as possible. Such a data-driven approach enables the learner to better capture the time-varying nature of risk preferences and better leverage the power and opportunities offered by ``Big data".

The applications of RL for investment management are not as plentiful as in other domains, such as games and robotics, partly due to the scarcity of the available data in the financial domain. Recently,  \cite{RL_MV} proposes a data-efficient RL algorithm to solve the MV problem in continuous time and space, while \cite{RL_MV2}  develops a more efficient way of using limited real data to train the RL algorithm in high-dimensional action space. Both works focus on training allocation strategies either without constraints or with gross leverage constraint. Our work handles the no-shorting, no-borrowing constraint using deep neural network approximations with a special structure for the input data. A similar investment optimization problem under Dirichlet policies has been recently studied in \cite{dirichlet}. It is interesting to note that their algorithm also generates equal-weight strategy in most cases, although the algorithm is trained using firm-level data rather than just market data.

\section{Conclusion} \label{sec:conclusion}
We have proposed a full-cycle, data-driven (model-free) investment robo-advising framework that leverages both inverse optimization and deep reinforcement learning techniques. Such an automated investment advising model can naturally lead to a personalized robo-advisor who can help the retail or institutional investors automatically determine their risk profiles and portfolio allocation decisions in real time. Our approach learns the unknown risk profile of an investor using a two-layer iterative scheme, capturing both the risk preference and target return. A downstream multi-period portfolio optimization problem is further formulated based on the learned risk profile. With deep reinforcement learning and quantitative analysis, we can obtain a dynamic rebalancing strategy catered for the investor . Our learned investment strategy, when applied to the overall market as a whole, has demonstrated superior performance against other comparative strategies over the period from April 1, 2016 to February 1, 2021.

Interesting future research may relate to other model-based robo-advising frameworks (e.g., \cite{capponi}). Our framework can also be useful to learn other types of investment strategies, for example, a dynamic index-tracking strategy if the investor prefers passive investment rather than active investment.

\bibliography{bibfile}

\clearpage

\appendix

\section{Implementation} \label{sec:data}
This section provides the details on the data collected by the IPO agent and the DRL agent. We also elaborate on how the data has been used in our experiments, in addition to the implementation details.
\subsection{Implementation for IPO agent}

\subsubsection{Price data and portfolio data}\label{section:price data and portfolio data}
We collect asset-level holdings of S\&P 500 portfolio from quarterly reports publicly available on the SEC website. Starting from March 2010, we collect portfolio $Y_t$ in each quarter $t$ till March 2020. The total number of observations is $T=40$. In a mutual fund portfolio, assets excluded or newly included in the 10-year duration are all included, and if at a specific time $t$ an asset does not appear in holding, its weight is 0. 

We query twenty years of historical asset-level price data from January 1, 2000 to March 31, 2020 using AlphaVantage API \cite{alphavantage}. The yearly profit of individual asset $p_{i,t}$ is calculated as a ratio. We use a moving window of 252 days (corresponding to the number of trading days per year) on the daily price sequence data. For each rolling window, the closing price at the last day of the window is the numerator, and the denominator is the closing price of the first trading day of the window. In this way we obtain the rolling yearly profit at day level. We then further aggregate this day-level rolling yearly profit to month level by averaging the profit values of all trading days in the same month. Thus, we denote by $\mathbf{p}_{t}$ the vector of yearly profit ratio, where each element corresponds to each individual asset at month $t$. We let $P_{t} = [\mathbf{p}_{1},...,\mathbf{p}_{t}]$.  

At each month $t$, we consider the backward history of $w$ months as the observation of market signals. The covariance matrix is calculated as
	\begin{align}
	Q_{t} & = \textrm{cov}([\mathbf{p}_{t-w},...,\mathbf{p}_{t}],[\mathbf{p}_{t-w},...,\mathbf{p}_{t}]).
	\end{align}
The initial expected asset-level return $\mathbf{c}_{0}$ is set to be the simple mean of historical profits during the observation period [$\mathbf{p}_{t-w},\dots,\mathbf{p}_{t}$]. The initial risk preference $r_{0}$ is usually set to a small constant value such as $1$ or $2$.
	
\subsubsection{Time-varying Observations}\label{section:Time-varying Observations}
\begin{figure}[ht]
\centering
	\includegraphics[width=.5\textwidth]{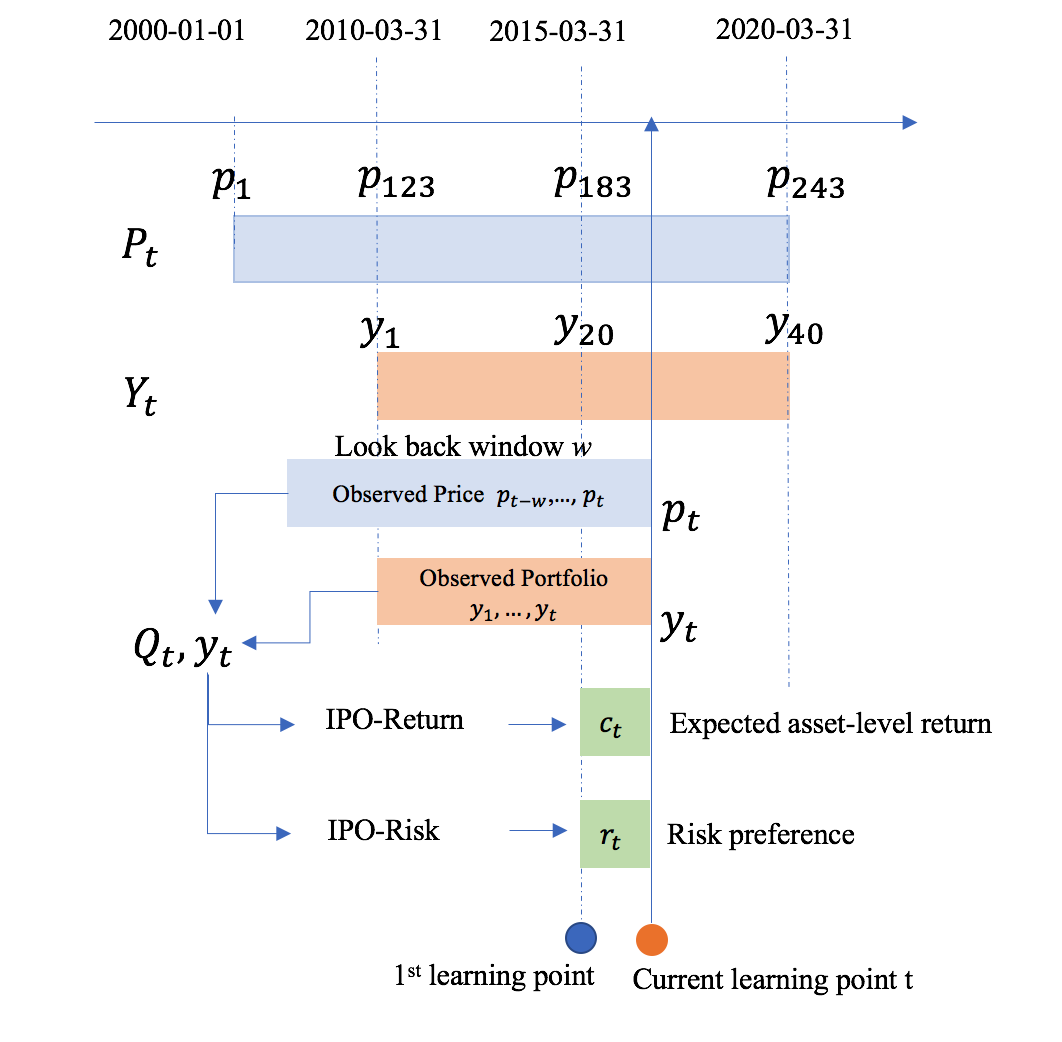}
	\caption{Time-varying observations for risk preference and expected return learning.}%
	\label{figure: time-varying observations_1}
\end{figure}

As illustrated in Figure \ref{figure: time-varying observations_1}, we align portfolio $Y_{t}$ and price $P_{t}$ on the same timeline. We start our learning from March 2015, which corresponds to $t=183$ for $P_{t}$ and $t=20$ for $Y_{t}$. At the first learning point, if we use the entire observed price history, then $w=t-1$. We use price data from January 2000 to December 2014 to estimate the covariance matrix $Q_{t}$ and the initial asset-level return guess $\mathbf{c}_{0}$. Then, we use historical portfolio observations $\mathbf{y}_{1},...,\mathbf{y}_{t}$ (starting from March 1, 2010) to learn $r_{t}, \mathbf{c}_{t}$. For the next learning point, we shift forward three months. The number of portfolio observations increases by 1 as $t$ increases by 1. For example, when $t=20$, we use $\mathbf{y}_{1},...,\mathbf{y}_{20}$ to estimate $r_{t}, \mathbf{c}_{t}$. When $t=21$, estimation is done by observing $\mathbf{y}_{1},...,\mathbf{y}_{21}$. So the more recent estimations of $r_{t}, \mathbf{c}_{t}$ have more observations of $\mathbf{y}_{t}$. When $t$ changes, the algorithms reset the initial guess of $r_{0}$ and $\mathbf{c}_{0}$, and we use the same initial guess $r_{0}$ at different $t$, and $\mathbf{c}_{0}$ is the simple average of the new sequence $[\mathbf{p}_{1},...,\mathbf{p}_{t}]$. In other words, at different $t$, the estimations obtained from previous observation time $t-1$, known as $r_{t-1}, \mathbf{c}_{t-1}$, have no impact on the current $r_{t}, \mathbf{c}_{t}$. Each time when $t$ changes, the learning restarts from the first element till the last element of the sequence ($\mathbf{y}_1,...,\mathbf{y}_t$) again.

\subsection{Implementation for DRL agent}
The DRL agent relies on historical daily price data of all the S\&P 500 constituent stocks to train, validate and test the DDPG algorithm. In particular, the test periods are from April 1, 2016 to February 1, 2021, separated into multiple one-year moving windows. 

For each one-year test period, we use the previous two-year daily data to train and validate the DDPG algorithm. Based on the validation using the second year price data, we choose the DDPG model that generates the highest average Sharpe ratio (among $100$ validating portfolios) as the best model, and its performance is then tested on the subsequent one-year testing data. For all the tests, we randomly select $100$ portfolios each consisting of $n$ stocks and provide investment outcomes regarding their performance, including the average wealth (i.e., portfolio value) process and the $95\%$ confidence bands as shown in Figure \ref{fig:figure1}. In addition, we also calculate the performance statistics, such as the cost-adjusted return, standard deviation, Sharpe ratio, cumulative traction cost, turnover rate and maximum drawdown, for the learned portfolio rebalancing strategies, provided in Table \ref{tab:table2}. All these statistics are based on the average results of $100$ randomly selected portfolios. Specifically, for a single portfolio, the daily turnover rate is computed using the following definition from \cite{equal}
\begin{equation}\label{turnover}
\text{Turnover rate} = \frac{1}{N-1}\sum_{i=0}^{N-2}\sum_{j=1}^n \Big|\frac{v^j_{i+1}-v^j_{i^+}}{W_{i+1}}\Big|\times 100\%.
\end{equation}
Here, $v^j_{i+1}$ is the dollar value to be invested in stock $j$ at the $(i+1)$-th rebalancing time, while $v^j_{i^+}$ is the holding value  in stock $j$ right {\it before} the $(i+1)$-th rebalancing happens. The transaction cost is related to the turnover (\ref{turnover}); it is generated by the accumulative turnover within the test period, using a proportional transaction cost rate $c$, i.e.,
\begin{equation}\label{transaction}
\text{Transaction cost}=\frac{c}{W_0}\sum_{i=0}^{N-2}\sum_{j=1}^n\big|v^j_{i+1}-v^j_{i^+}\big|\times 100\%.
\end{equation}
This formula computes the total cost incurred from each transaction order (including both buy and sell orders) within the investment horizon during test. To comply with recent empirical studies (see, e.g., \cite{trading_cost}), we choose $c$ to be $5$ basis points per transaction of the underlying S\&P 500 stocks in our experiments. Finally, the maximum drawdown is defined as the maximum observed loss from a peak to a trough of a portfolio, before a new peak is attained, i.e., 
\begin{equation}\label{MD}
\text{Maximum drawdown} = \max_{0\leq i_1\leq i_2\leq N-1} \frac{W_{i_1}-W_{i_2}}{W_{i_1}},
\end{equation}
where $W_i$ is the portfolio value at date $i$.

\end{document}